\documentstyle{article}

\begin{document}

\title{A Rational Approach to  Cryptographic Protocols}
\date{}
\author{P. Caballero-Gil, C. Hern\'{a}ndez-Goya and C. Bruno-Casta\~neda\\
{\small Department of Statistics, Operations Research and Computing,}\\
{\small Faculty of Mathematics, University of La Laguna,}\\ 
{\small 38271 Tenerife, Spain}
\\
{\small Corresponding author: pcaballe@ull.es }}
\maketitle
\begin{abstract}
This work initiates an analysis of several cryptographic protocols
from a rational point of view using a game-theoretical approach,
which allows us to represent not only the protocols but also
possible misbehaviours of  parties. Concretely, several concepts
of two-person games and of two-party cryptographic protocols are
here combined in order to model the latters as the formers. One of
the main advantages of analysing a cryptographic protocol in the
game-theory setting is the possibility of describing improved and
stronger cryptographic solutions because possible adversarial
behaviours may be taken into account directly. With those tools,
protocols can be studied in a malicious model in order to find
equilibrium conditions that make possible to protect honest
parties against all possible strategies of adversaries.

\end{abstract}
\section{Introduction}
\footnotetext{Research partially supported by the Spanish Ministry of Education
and Science and the European FEDER Fund under SEG2004-04352-C04-03
Project.\\
Mathematical and Computer Modelling. Volume 46, Issues 1-2, July 2007, Pages 80-87. \\ 
DOI:10.1016/j.mcm.2006.12.013  
}
The verification of cryptographic protocols has become a subject
of great importance with the development of communications and
transactions on public channels like Internet. Since Cryptology
may be seen as a continuous struggle between cryptographers and
cryptanalysts, and Game Theory may be defined as the study of
decision making in difficult situations, both fields seem to have
certain common scenarios, so it is natural that tools from one
area may be applied in the other. In fact, the main objective of
this work is to model several two-party cryptographic protocols as
two-person games in order to introduce the human factor in the
analysis of cryptographic protocols so that it might be helpful to
solve many security problems which are hard to deal with
traditional security primitives.

One of the first approaches that analyses the relationship between
cryptographic protocols and games may be found in
\cite{Fischer-Wright93}, where
 an application of game theoretic techniques to the
analysis of some multiparty cryptographic protocols for secret
exchange was provided. Later, a solution to the problem of
determining the existence of two-person games whose payoffs are
comparable to those obtained when a Third Trusted Party intervenes
was proposed in \cite{Dod:Hal:Rab00}. Another two recent
applications of modern cryptography to game theory were presented
respectively in \cite{Gos99}, where it was proved that every
correlated equilibrium of an original infinitely repeated game can
be implemented through public communication only, and in
\cite{BS04}, where cryptographic primitives were used to provide
correctness and privacy in distributed mechanisms.

Several cryptographic proofs of protocols correctness based on
basic fairness were provided in \cite{GarayJakobssonMacKenzie99},
whereas in \cite{BuH00} various formal definitions of different
versions of fairness were given. The idea of using game theory as
a formal tool to model specific cryptographic protocols such as
Fair and Safe Exchange, and Contract Signing was explored in the
recent works \cite{Kre:Ras02}, \cite{SW02} and \cite{CMSS03}. The
concept of rational exchange in terms of Nash equilibrium was
defined in \cite{ButtyanHubaux01}, where it was proved that fair
exchange implies rational exchange but not the reverse. Another
remarkable reference, \cite{AsokanShoupWaidner00}, described a
formal security model for fair signature exchange in terms of
games where fairness was defined in a probabilistic way.

Finally,  the work \cite{ButtyanThesis}  should be singled out as
the main starting point of this work since there the concept of
rationality applied to exchange was introduced. Such a reference
also showed the close relationship between the rationality concept
and the stimulation for cooperation in ad-hoc networks.

This paper represents a preliminary step of a game-based analysis
of general scenarios and different types of two-party
cryptographic protocols. Concretely, here the modelling of
incentives in the games and desirable conditions of the protocols
are described. The structure of the present work is as follows.
Section 2 introduces briefly notations and definitions of several
game theoretic notions that are used throughout the paper. Then
Section 3 provides a basic background on two-party cryptographic
protocols. In Sections 4 and 5 a theoretic game model is used to
describe and analyse respectively symmetric and asymmetric
two-party protocols. Finally, conclusions of the work and comments
on further investigation are drawn in Section 6.

\section{Notations and Definitions}

If a group $P$ of parties or players $i$ agree to obey certain
rules and to act individually or in coalition, the results of
their joint action lead to certain situations called outcomes. In
such conditions, a game $G$ defines the set of rules that specify
a sequence of actions $a \epsilon  Q$ allowed to the parties.

Concretely, the rules of the game specify what amount of
information about all the previous actions and the alternatives
that have been chosen can be given to each party before making an
specific choice. The game also specifies a termination when some
specific sequences of choices are made and no more actions are
allowed. Each termination produces an outcome in the form of
scores or incomes $y^+$, and payments or expenses $y^-$ for each
party. It is assumed that each party $i$ has a preference relation
$\leq _i$ over the outcomes reflected in his/her scores and
payments.

A finite action sequence $q$ is said to be terminal if it is
infinite or if there is no action $a$ such that $q$ is followed by
$a$. The set $Z$ of terminal action sequences represents all the
possible outcomes of the game. The real-valued  function
$y(q)=(y_i (q))_{i \epsilon  P}$ that assigns the payoffs for
every party $i$ after every terminal action sequence $q \epsilon
Z$ is called outcome or payoff function. These payoff values may
be negative, in which case they are interpreted as losses. Also
these payoffs may verify that $\sum_{i\in P} y_{i}(q) = 0$ for any
$q \in Z$, in which case the game is called zero-sum.

The preference relations of the parties are often represented in terms of their payoffs in
such a way that for any $q,q' \epsilon  Z$ and $i \epsilon  P$, $q\leq_i q'$ iff $y_i(q)\leq
y_i(q')$. On the other hand, the so called utility function $u_i$ is just a mathematical
representation of $i$'s preferences.

A strategy of party $i \epsilon  P$ is a function $s_i \epsilon
S_i$ that assigns an action which is available after $q$ for party
$i$, to every non-terminal action sequence $q \epsilon Q\backslash
Z$ such that $i$ is the following party in choosing an action
after $q$. A strategy profile is a vector $(s_i)_{i \in P}$ of
strategies, where each $s_i$ is a member of $S_i$. The notation
$(s_j,(s_i)_{i \epsilon P\backslash{j}})$ is used to emphasise
that the strategy profile specifies strategy $s_j$ for party $j$.
Finally, let $o((s_i)_{i \epsilon P})$ denote the resulting
outcome when the parties follow the strategies in the strategy
profile $(s_i)_{i \epsilon P}$.

A strategy profile $({s^*}_i)_{i \epsilon P}$ is called a Nash
equilibrium iff for every party $j \epsilon P$ we have that
$o(s_j,(s^*_i)_{i \epsilon P\backslash{j}})\leq_j o(s^*_j,
(s^*_i)_{i \epsilon P\backslash{j}})$. This means that if every
party $i$ other than $j$ follows strategy $s^*_i$, then party $j$
is also motivated to follow strategy $s^*_j$. So, in Nash
equilibrium  the choices depend on the other's possible
strategies.

\section{Cryptographic Protocols Concepts}

A two-party cryptographic protocol may be defined as the
specification of an agreed set of rules on the computations and
communications that need to be performed by two entities, $A$
(Alice) and $B$ (Bob), over a communication network, in order to
accomplish some mutually desirable goal, which is usually
something more than simple secrecy. Several essential properties
of cryptographic protocols are the following:
\begin{enumerate}
\item Correctness, which guarantees that every honest party should get his/her agreed output.

\item Privacy, which includes the protection of every party' secrets.

\item Fairness, which means that if a dishonest party exists, then neither he/she may gain
anything valuable,  nor honest party may lose anything valuable.

\end{enumerate}

In the game-theoretic model two new properties regarding dishonest
behaviours can be defined

\begin{enumerate}
\item Exclusiveness, which implies that one or both parties cannot receive their agreed output.

\item Voyeurism, which is the contrary of privacy because it implies that one or both parties may discover the other's secret.
\end{enumerate}

Note that the previous definition of fairness agrees with the
rationality concept described in \cite{ButtyanHubaux01} because
fairness here is a property which is understood more practical
than theoretical. In other words, protocols are here defined
according to their practical security against any kind of
adversaries.

It is assumed that at each step a party receives the message that
was sent by the other  party at the previous step, performs some
private computation and sends some message (possibly none) to the
other party. So, a two-party cryptographic protocol may be seen as
a repeated game formed by a sequence of iterations of the
following two communication phases:

1)Send: Party $A$ ($B$) sends to $B$ ($A$) a message $M$ generated
depending on her (his) state.

2)Receive: Party $A$ ($B$) receives from $B$ ($A$) a message $M$
and makes a state transition.

Thus, we are implicitly assuming that the system is synchronous
(parties know the time and must decide what message to send in
each round before receiving any message sent to them in that
round), communication is guaranteed, and messages take exactly one
round to arrive. These assumptions are critical to the correctness
of the protocols. Also, for the sake of simplicity, in this paper
 the non-intentional loss of control over message $M$ is considered as
a delivery, so  $rcv_A(M)$ ($rcv_B(M)$) denotes both the cases
when party $B$($A$) sends message $M$ to $A$($B$), and when
$A$($B$) is able to receive it.

In order to formalise the notion of cryptographic protocols in
terms of functions, we denote by  $f$ a two-argument finite
function, $f: X_A \times X_B \rightarrow Y_A \times Y_B$ where
$X_i$ and $Y_i$,  $i \epsilon \{A,B \}$, represent respectively
input and output sets for party $i$. Intuitively, a two-party
cryptographic protocol may be generally described through a
two-variable function $f$ whose output is defined by  the
expression $f(M_A,M_B)=(f_{A}(M_A,M_B),f_{B}(M_A,M_B))$, where it
is understood that party $i$ receives the output of $f_i$ on
inputs $M_A$ and $M_B$.

As aforementioned, two-party cryptographic protocols include a
series of message exchanges between both parties over a
communication network. So, the possibility always exists that one
or both parties will cheat to gain some advantage or that some
external agent will interfere with normal communications. The
simplest situation occurs when each party functions asynchronously
from the other party and makes inferences by combining a priori
knowledge with properties of the received messages, determining
information that is not immediately apparent, so such inferences
must be taken into account in determining security. In a worst
case analysis of a protocol, one must assume that any party may
try to subvert the protocol. So, when designing a two-party
cryptographic protocol one of two possible models should be
considered:

\begin{itemize}
\item Semi-honest model: When it is assumed that the protocol is cooperative and
 both parties follow the protocol properly in such a way that they help each other to compute $f_{i}(M_A,M_B)$, but
curious parties may keep a record of all the information received
during the execution and use it to make a later attack.

\item Malicious model: Where it is assumed that parties may deviate from the protocol. In this
case, during the interaction, each party acts non cooperatively
and has different choices which may determine the output of the
protocol.
\end{itemize}

We are interested in obtaining guarantees provided by the
definition of the protocols when one of both parties misbehaves in
an arbitrary way. Consequently, this work is conducted within the
malicious model where it is assumed that either $A$ or $B$ does
not follow the protocol properly. In such a model the security of
a cryptographic protocol should refer to its ability to withstand
attacks by certain types of cheaters or enemies, in such a way
that essential properties such as correctness, privacy and
fairness hold despite such possible attacks. So, the main interest
of this work will be the description of honest strategy profiles
for every analysed protocol such that whenever the strategy of
some party is honest, the other party has no incentive to deviate
from the protocol, which is closely related to Nash equilibrium
conditions.

Apparently, any two-party cryptographic protocols might be best
modelled with a zero-sum game because every situation that is
dishonestly advantageous for a party should be disadvantageous for
the other. In fact this is not the case of many protocols. In
general, most two-party cryptographic protocols are represented by
non-positive sum games (i.e. games in which the sum of the payoffs
of the players is always less than or equal to 0). Those games in
which the sum of the payoffs can be positive should be generally
discarded because they imply that both parties could agree on
behaving dishonestly and receive positive payoffs.

In particular, the payoff  $y_i(q)$ of a party $i$, assigned after
a terminal action sequence $q$ may defined as
$y_i(q)={y^+}_i(q)-{y^-}_i(q)$, where ${y^+}_i(q)$ and
${y^-}_i(q)$ represent respectively the incomes  and expenses of
$i$ after $q$. These incomes and expenses functions will be
defined in terms of utilities according to the concrete
definitions of each protocol. Here the  utility that a secret
$M_j$ is worth to party $i$ is denoted by $u_{ij}=u_i(M_{j})$,
value which may be difficult to quantify in practical situations.

A two-party cryptographic protocol is said to be closed when if a
party gains something, then the other party must lose something.
This property may be expressed in terms of the incomes and
expenses functions in the following way: $\forall q \epsilon Z,
{y^+}_i(q)>0 \Rightarrow {y^-}_j(q)>0$. Note that in this work the
closeness of the protocols is assumed since in the definition of
the payoff function we always consider both the wish of one party
to know the other's secret and the wish of the other party to
prevent that from happening.

According to the aforementioned functional definition of a
two-party cryptographic protocol $f$, at the end of the execution,
party $i$ should receive the output of $f_i$ on secrets $M_A$ and
$M_B$. Depending on whether $f_A=f_B$ we may distinguish between
symmetric and asymmetric protocols. From the first group, in the
next sections we will study the protocols of Fair Exchange, Secure
Two-Party Computation and Coin Flipping. On the other hand,
representative protocols of the group of asymmetric protocols are
Oblivious Transfer, Bit Commitment and Zero Knowledge Proof. This
classification is important for the proposed game theoretic model
because it implies the translation to a symmetric game where
possible payoffs and outputs of both parties coincide, or to
asymmetric games where that does not occur.

In the following sections several symmetric and asymmetric
protocols are analysed according to a game-theoretic model. For
every analysed protocol we define income, expense and payoff
functions for each party in every possible combination of
behaviours and misbehaviours of parties, and make rather minimal
assumptions about several matters such as the preferences of the
parties in order to guarantee the existence of a honest strategy
profile being a Nash equilibrium. Although the possibility of
misbehaviours by both parties is here considered, in this paper we
analyse specially the case when exactly one of them is dishonest.
Note that if this assumption is not fulfilled, there might be some
dishonest strategy that dominates the corresponding honest
strategy, and in such conditions rational parties would be
consequently dishonest.

\section{Symmetric Protocols}

\subsection{Fair Exchange}

Fair Exchange is a  cryptographic protocol for exchanging secrets
$M_A$ and $M_B$ between two parties $A$ and $B$ so that if $A$
behaves correctly, then party $B$ cannot get $A$'s secret $(M_A)$
unless $A$ gets $B$'s secret $(M_B)$, and vice versa. According to
this definition, possible descriptions of non-null values of the
incomes and expenses functions ${y^+}_i$ and ${y^-}_i$  are the
following:

${y^+}_i(q)=u_{ij}$ if $rcv_i(M_j)$

${y^-}_i(q)=u_{ii}$ if $rcv_j(M_i)$.

Note that if no assumptions or preferences of parties are made,
rational parties will simply not send their secrets since this
strategy weakly dominates sending the secret. However, since the
parties' objective in this protocol is to obtain each other's
secret, we are only interested in the states of the protocol tree
where $A$ possesses $B$'s secret and the ones where $B$ possesses
$A$'s secret. So, one property that utility $u_{ij}$ should verify
in order to avoid a possible coalition between two dishonest
parties is the following: $u_{ij} > u_{ii} > 0, \forall
i,j\epsilon \{A,B\}$.  For example, such utilities might reflect
the interests of both parties to participate cooperatively if the
protocol is run correctly. In this way, parties value correctness
over privacy, and the payoff $y_i(q)$ of party $i$ can take only
four possible values: $-u_{ii} < 0 < u_{ij}-u_{ii} < u_{ij}$
corresponding respectively to the four possible terminal action
sequence when $rcv_j(M_i) \leq_i rcv_i(\emptyset) \wedge
rcv_j(\emptyset) \leq_i rcv_i(M_j) \wedge rcv_j(M_i) \leq_i
rcv_i(M_j)$.

Fairness property ensures that if $i$ is honest, then the other
party $j$ cannot get $i$'s secret unless $i$ gets $j$'s secret.
So, in terms of incomes and expenses functions we have that if
$i$'s strategy $s^*_i$ is honest, then for every strategy of $j,
s_j$:
 if $y^+_j(o(s^*_i,s_j))=u_{ji} \Rightarrow
y^+_i(o(s^*_i,s_j))=u_{ij}$.

 So, it may
be stated that in a rational fair exchange protocol where both
parties have incentives to send their secrets, honest strategies
are Nash equilibrium because if one party follows a honest
strategy, then the other party is also motivated to behave
honestly because he/she loses or at least does not gain anything
by not doing so.

Examples of fair exchange include Contract Signing and Certified
Mail protocols \cite{EGL85}. In the former, both parties $A$ and
$B$ want to exchange simultaneously signed contracts in such a way
that none of them can obtain the signature of the other without
having signed the contract and that none of them can repudiate his
or her own signature. On the other hand, in Certified Mail $A$
wants to send a mail $M_A$ to $B$ so that $B$ can read the mail
$M_A$ if and only if $A$ receives the corresponding return receipt
$M_B$. Consequently, a  conclusion similar to the obtained for
fair exchange may be extracted for both cases of contract signing
and certified mail protocols.

\subsection{Secure Two-Party Computation}

The general protocol known as Secure Two-Party Computation allows
that two parties $A$ and $B$ with secret inputs $M_A$ and $M_B$ to
evaluate a common value $f_A(M_A,M_B)=f_B(M_A,M_B))=g(M_A,M_B)=g$
 in a manner where neither party learns more than necessary.
 This protocol is the two-party version of
the multiparty protocol known as Secure Function Evaluation. There
 are various definitions and models for Secure Two-Party
 Computation \cite{Gol02} and indeed the above definition describes just one of
 them. For example, one might consider an asymmetric version where
 only $A$ receives the output. However,  this work deals with this symmetric version where both parties learn the value $g$.

A possible description of the incomes and expenses functions
${y^+}_i$ and ${y^-}_i$ that verifies the previous definition is
as follows, where $k>1$:

${y^+}_i(q)=  \left\{\begin{array}{ll}
  u_{ij} & if \ \  rcv_i(M_j) \\
  k u_i(g) & if  \ \  rcv_i(g) \\
  u_{ij}+k u_i(g)&  if\ \ rcv_i(M_j,g) \\
  0 & otherwise
\end{array}\right.$

$
{y^-}_i(q)= \left\{\begin{array}{ll}
  u_{ii} & if \ \ rcv_j(M_i)\\
  u_i(g)) & if \ \  rcv_j(g) \\
  u_{ii}+u_i(g)&  if \ \ rcv_j(M_i, g) \\
  0 & otherwise
\end{array}\right .$

A serious problem of this protocol arises when there is no way to
force a party to use his/her correct input. So, according to
privacy property, and in order to avoid a possible coalition
between dishonest parties, we there should be assumed that the
following inequality holds: $u_i(g)< u_{ij}< k u_i(g)< u_{ii},
\forall i,j\epsilon{A,B}$ which implies that: exclusiveness
$\leq_i$ voyeurism $\leq_i$ correctness $\leq_i$ privacy.

If the utility of $g(M_A,M_B)$ is the same for both parties,
$u=u_A(g)=u_B(g)$, the payoff $y_i(q)$ of party $i$ may take the
following sixteen possible values: $-u-u_{ii}<
 -u_{ii} <
 u_{ij}-u_{ii}-u <
 (k-1)u - u{ii}<
ku-u_{ii}
 < -u,
u_{ij}-u_{ii}, (k-1)u+u_{ij}-u_{ii} < 0, ku+u_{ij}-u_{ii}  <
u_{ij}-u
  < u_{ij},(k-1)u <
  ku <
  (k-1)u + u{ij} <
  ku+ u_{ij}$
corresponding respectively to the sixteen possible terminal action
sequence when $rcv_j(g,M_i) \leq_i rcv_j(M_i) \leq_i rcv_j(g,M_i)
\wedge rcv_i(M_j) \leq_i rcv_i(g) \wedge $ $ rcv_j(g, M_i) \leq_i
rcv_j(M_i) \wedge rcv_i(g) \leq_i rcv_j(g), rcv_j(M_i) \wedge
rcv_i(M_j), $ $ rcv_j(g,M_i) \wedge rcv_i(g,M_j) \leq_i
rcv_i(\emptyset) \wedge rcv_j(\emptyset),$
 $rcv_i(g, M_j)
\wedge rcv_j(M_i) \leq_i
$
$ rcv_i(M_j) \wedge rcv_j(g)  \leq_i rcv_i(M_j), rcv_i(g) \wedge
rcv_j(g)$
$ \leq_i rcv_i(g) \leq_i $ $ rcv_i(g, M_j) \wedge
rcv_j(g) \leq_i rcv_i(g, M_j)$.

A rational secure two-party computation protocol ensures that no
party receive the other party's secret and that if party $i$ is
honest, then the other party $j$ cannot get $g(M_A, M_B)$ unless
$i$ gets it. So, in terms of incomes and expenses functions we
have that if $i$'s strategy $s^*_i$ is honest, then for every
strategy of $j$, $s_j$: if $y^+_j(o(s^*_i,s_j))=k u_j(g)
\Rightarrow y^+_i(o(s^*_i,s_j))=k u_i(g)$, and if
$y^+_j(o(s^*_i,s_j))=u_{ji}+k u_j(g) \Rightarrow
 y^+_i(o(s^*_i,s_j))=u_{ij}+ku_i(g)$.
So,  in  rational secure two-party computation protocol, honest
strategies hold Nash equilibrium conditions.

\subsection{Coin Flipping}
Coin flipping protocols are used where two parties $A$ and $B$
want to generate jointly a common random binary sequence $M$.
According to this definition, possible descriptions of non-null
additive values of the incomes and expenses functions ${y^+}_i$
and ${y^-}_i$  are the following, where $k>1$:

${y^+}_i(q)=u_i(M)$ if $M$ is selected by $i$

${y^+}_i(q)=ku_i(M)$ if $rcv_i(M)$

${y^-}_i(q)=k u_i(M)$ if $M$ is selected by $j$

${y^-}_i(q)=u_i(M)$ if $rcv_j(M)$.

In this way,  according to preferences of parties, correctness and
voyeurism are valued over exclusiveness and privacy, and the
payoff $y_i(q)$ of party $i$ can take five possible values: $-k
u_i(M) < -u_i(M)< 0 < u_i(M)< k u_i(M)$ corresponding respectively
to the five possible terminal action sequence when $M$ is selected
by $j \leq_i rcv_j(M) \leq_i rcv_j(M) \wedge rcv_i(M) \leq_i M$ is
selected by $i \leq_i  rcv_i(M)$

Again fairness property ensures that either both parties get the
agreed outcome or neither does, so if party $i$ is honest, then
the other party $j$ cannot get the randomly generated sequence $M$
before. So, in terms of incomes and expenses functions we have
that if $i$'s strategy $s^*_i$ is honest, then for every strategy
of $j, s_j$, if $y^+_j(o(s^*_i,s_j))=ku_j(M) \Rightarrow
y^+_i(o(s^*_i,s_j))=ku_i(M)$. Consequently,  honest strategies in
rational coin flipping protocols are Nash equilibrium.

\section{Asymmetric Protocols}

\subsection{Oblivious Transfer}

A major component in the construction of Secure Two-Party
 Computation protocols is the  Oblivious Transfer protocol since
 it has been proved that a Secure Two-Party
 Computation can be always built using calls to an Oblivious Transfer
 protocol \cite{Rab81}. So, the term Oblivious Transfer refers usually to
 several different versions of asymmetric Secure Two-Party
 Computation protocols, all of which turned out to be equivalent.
 However, the definition that will be used in this work is the
 following. An Oblivious Transfer may be defined as a protocol whose goal is to enable one
party $A$ to transfer a secret to another party $B$ in such a way
that the information is transferred with a probability 1/2, and
when concluding the protocol $B$ knows with absolute certainty
whether he has got the secret or not, but $A$ does not know it.

Possible descriptions of additive incomes and expenses functions
${y^+}_i$ and ${y^-}_i$ are the following, where $k>1$:

${y^-}_A(q)=u_A(M)$ and ${y^+}_B(q)=u_B(M)$ if $rcv_B(M)$

${y^+}_A(q)=k u_A(M)$ and ${y^-}_B(M)=(k+1)u_B(M)$ if $A$ knows
whether $rcv_B(M)$ or not.

If no assumption is made on $A$' interest to participate in a
correct protocol, then there may be a problem because a rational
party $A$ would simply not send her secret. Consequently, the
described model implies that party $A$ should value voyeurism over
exclusiveness, whereas party $B$ should value privacy over
correctness.  On the one hand, the payoff functions of party $A$
can take the following four values: $-u_A(M)<
(k-1)u_A(M)<0<ku_A(M)$ corresponding respectively to the four
possible terminal action sequence when $rcv_B(M) \leq_i rcv_B(M)
\wedge A$ knows it $ \leq_i rcv_B(\emptyset) \leq_i
rcv_B(\emptyset)\wedge A$ knows it. On the other hand, the payoff
functions of party $B$ can take the following four values:
$(-k-1)u_B(M)< -ku_B(M)<0<u_B(M)$ corresponding respectively to
the four possible terminal action sequence when
$rcv_B(\emptyset)\wedge A$ knows it $\leq_i rcv_B(M) \wedge A$
knows it $ \leq_i rcv_B(\emptyset) \leq_i rcv_B(M)$.

A rational oblivious transfer ensures that if party $B$ is honest,
$A$ cannot know whether $B$ received the secret or not, and if
party $A$ is honest, $B$ receives the secret with probability 1/2.
So, in terms of incomes and expenses functions we have that if
$B$'s strategy $s^*_B$ is honest, then for every strategy of $A$,
$s_A$: if $y^+_A(o(s^*_B,s_A))=k u_A(M) \Rightarrow
y^+_B(o(s^*_B,s_A))=u_B(M)$, so  $s_A$ is not a good strategy for
$A$. From the above it may be stated that  honest strategies in
rational oblivious transfer hold Nash equilibrium conditions.

\subsection{Bit Commitment}

The goal pursued by this two party protocol is twofold: first $A$
transfers information to $B$ that can not be changed for her
(unalterability property) and such information can not be accessed
by $B$ until the end of the protocol is reached (illegibility
property). Originally the aforementioned information consists of
only one bit.

When defining utility function the possible frauds should be taken
into account for both participants. So, in this case $B$ would
obtain the bit before opening the commitment, $A$ could also
modify the content of the original commitment while the protocol's
development. The expenses and incomes of each participants are the
following where $k>1$:

$ y_A^-(q) = k \cdot u_A(M), y_B^+ = u_B(M)$, if $rcv_B(M))$
before the opening stage

$y_A^+(q) = u_A(M), y_B^- = k \cdot u_B(M)$, if $A$ modifies $M$

 According
to the previous values, the payoff for each party has the values
0, $-k \cdot u_i(M), u_i(M)$ and $(1-k)\cdot u_i(M), i \in
\{A,B\}$. From these utility functions it can be deduced that the
honest behaviour of party $A$ implies $B$'honesty, since the other
possibilities convey non positive payoffs. Again honest strategies
have Nash equilibrium associated. Furthermore it can be deduced
that party $A$ associate a bigger weight to privacy property than
to exclusiveness. On the other hand, $B$'s preferences single out
correctness property compared to voyeurism.

\subsection{Zero-Knowledge Proofs} A zero-knowledge protocol allows
party $A$ to convince $B$ that she knows some information but
without leaking anything about the secret. The two dishonest
possibilities considered are: party $A$ does not know the secret
or party $B$ gets the secret, so the corresponding expenses and
incomes are

 $ y_A^- = k \cdot u_A(M), y_B^+ =
u_B(M)$, if $rcv_B(M)$

$y_A^+ = u_A(M), y_B^- = k \cdot u_B(M)$ when $A$ does not know
the secret

The payoff deduced from those values are $0, u_i(M)$ and $-k \cdot
u_i(M), i \in \{A,B\}$. Hence, Nash's equilibrium forces both
participants to be honest. According to the previous model, party
$A$ should value privacy over exclusiveness while for party $B$,
correctness outweighs voyeurism.

\section{Conclusions}

This paper addresses an emergent issue in security: the synergy
between security protocols and game theory mechanisms. In
particular, the study of  several two-party protocols in a
game-theoretic model is here initiated, by giving formal
definitions of payoffs for each party and ranking properties of
exclusiveness, voyeurism, correctness and privacy. This work deals
with the idea of modelling cryptographic protocols design as the
search of an equilibrium in order to defend honest parties against
all possible strategies of malicious parties. So, our first
objective has been to illustrate the close connection between
protocols and games and to use game theoretic techniques for the
definition and analysis of cryptographic protocols so that this
model might be used to build more effective and efficient security
protocols.

Two subjects that are being object of work in progress are the
generalization of the game-theoretic approach followed in this
work to multiparty cryptographic protocols, and the analysis of
the relationship between properties like fairness and different
game theoretic concepts, such as dominant strategic equilibrium.
Finally, one direction for further investigation involves the
study of the possibility of describing two-party protocols as
sequential games instead of repeated games, which might be more
convenient in many cases.

\end{document}